\documentclass[twocolumn,showpacs,preprintnumbers,amsmath,amssymb]{revtex4}

\usepackage{amsmath,amssymb,graphics,epsfig,subfigure,color,hyperref}
\usepackage{color}

\newcommand {\nn}    {\nonumber}

\begin{document}

\title{Thick brane in mimetic $f(T)$ gravity}

\author{Wen-Di Guo$^a$\footnote{guowd16@lzu.edu.cn}}
\author{Yi Zhong$^a$\footnote{zhongy13@lzu.edu.cn}}
\author{Ke Yang$^b$\footnote{keyang@swu.edu.cn}}
\author{Tao-Tao Sui$^a$\footnote{suitt14@lzu.edu.cn}}
\author{Yu-Xiao Liu$^a$$^c$\footnote{liuyx@lzu.edu.cn, corresponding author}}

\affiliation{$^{a}$Institute of Theoretical Physics $\&$ Research Center of Gravitation, Lanzhou University, Lanzhou 730000, China\\
             $^{b}$School of Physical Science and Technology, Southwest University, Chongqing 400715, China\\
             $^{c}$Key Laboratory for Magnetism and Magnetic of the Ministry of Education, Lanzhou University, Lanzhou 730000, China}

\begin{abstract}
We apply the mimetic $f(T)$ theory into the thick brane model. We take the Lagrange multiplier formulation of the action and get the corresponding field equations of motion. We find solutions for different kinds of $f(T)$. Besides, we investigate the stability of the mimetic $f(T)$ brane by considering the tensor perturbations of the vielbein. Localization problem is also studied and it is shown that the four-dimensional gravity can be recovered for all the solutions. The effects of the torsion show that for the polynomial form of $f(T)$, the zero mode has a split compared with that of $f(T)=T$, but the situations for the exponential form of $f(T)$ are similar to that of $f(T)=T$.
\end{abstract}



\maketitle

\section{Introduction}
The dark matter problem has been one of the most important issues in recent cosmology~\cite{Ade:2015xua,Alam:2016hwk}. However, the nature of dark matter is still a mystery, because it has never been observed directly. Theoretical physicists have been working hard to reveal the mask of dark matter. Modifying Einstein's general relativity (GR) is an important approach. Among the various modified gravities, mimetic gravity~\cite{Chamseddine:2013kea}, a Weyl-symmetric extension of GR, has attracted more and more attention recently. In mimetic gravity, the conformal degree of freedom is isolated by parameterizing the physical metric in terms of an auxiliary metric and a scalar field (the so-called mimetic scalar field). In this way, this scalar field becomes dynamical and can mimic the cold dark matter~\cite{Chamseddine:2013kea,Chamseddine:2014vna}. Besides, the late-time acceleration and inflation can also be explained in the extended mimetic $f(R)$ gravity~\cite{Nojiri:2014zqa,Momeni:2015gka,Odintsov:2015cwa}. It was pointed out in Ref.~\cite{Barvinsky:2013mea} that this theory is free of ghost instability for a  positive energy density of the mimetic fluid in cosmology. But the ghost degrees of freedom do not vanish in mimetic theory with higher derivatives~\cite{Ramazanov:2016xhp}. The Lagrange multiplier method was applied to this theory in Refs.~\cite{Chamseddine:2014vna,Barvinsky:2013mea,Golovnev:2013jxa,Chamseddine:2016uyr}, which can give an equivalent formulation to mimetic gravity. Connections between mimetic gravity and Einstein-aether theories were discussed in Refs.~\cite{Speranza:2015sta,Jacobson:2015mra}. For more details of mimetic gravity, see Ref.~\cite{Sebastiani:2016ras} and references therein.

Teleparallel equivalence of general relativity (TEGR) was first proposed to unify the gravity and electromagnetism by Einstein in 1928~\cite{Einstein}. It is equivalent to GR due to the fact that the difference between the torsion scalar and the Ricci scalar is only a boundary term. It provides a possible way to interpret gravity as a gauge theory~\cite{Aldrovandi:2013wha}. The background manifold of TEGR is not the torsion-free Riemannian spacetime but the curvature-free Weitzenb\"{o}ck geometry. And the dynamical fields are vielbein which are defined on the tangent space of any point in the spacetime. Inspired by the $f(R)$ gravity,  Bengochea and Ferraro generalized TEGR to $f(T)$ gravity and explained the acceleration of the universe~\cite{Bengochea:2008gz}. After that, $f(T)$ gravity was investigated widely~\cite{Wei:2011jw,Nunes:2016qyp,Ferraro:2018tpu,Xu:2018npu}, and the degrees of freedom as well as local Lorentz invariance were studied in Refs.~\cite{Li:2011rn} and~\cite{Li:2010cg}, respectively. More details of $f(T)$ gravity can be seen in Ref.~\cite{Cai:2015emx} and the references therein. Recently, Mirza and Oboudiat applied the mimetic method to $f(T)$ gravity~\cite{Mirza:2017afs}. They kept the vielbein being unchanged but performed the conformal transformation on the Minkowski metric of the tangent space. The Lagrange multiplier formulation of mimetic $f(T)$ theory was also given~\cite{Mirza:2017afs}. Besides, five fixed points which representing inflation, radiation, matter, mimetic dark matter, and dark energy dominated eras respectively were also found in this theory if some conditions are satisfied.

\begin{figure}[htb]
\begin{center}
\includegraphics[width=4cm]{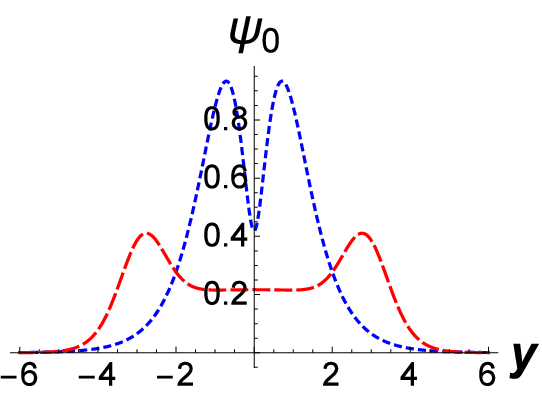}
\end{center}
\caption{Plots of the zero modes for different models. The blue dashed line is the zero mode for $f(T)$-brane generated by a matter field~\cite{Yang:2012hu} and the red dashed line with a cat-like shape is the zero mode for $f(T)$-brane generated by a mimetic field considered in this paper.}\label{figZeroModes}
\end{figure}

On the other hand, the extra-dimensional theory has been proposed since the 1920s~\cite{Kaluza:1921tu,Klein:1926tv}. And it attracted wide attention because of the well-known large extra dimension models~\cite{ArkaniHamed:1998rs,Antoniadis:1998ig,ArkaniHamed:1998nn} and warped extra dimension models (also called Randall-Sundrum (RS) models)~\cite{Randall:1999ee,Randall:1999vf}, which aimed at solving the hierarchy problem in the standard model of particle physics. After that, the thick brane model was proposed by combining the RS-2 brane model~\cite{Randall:1999vf} and domain wall model~\cite{Rubakov:1983bb,Rubakov:1983bz}.
More details on the thick brane and the localization of bulk fields on the brane were presented in the recent reviews~\cite{Liu:2017gcn} and references therein. The thick braneworld model in $f(T)$ theory was firstly constructed in Ref.~\cite{Yang:2012hu}. The authors gave the brane world solution for $f(T)=T+\alpha T^2$ and investigated the localization of bulk fermions. Besides, the split of brane caused by the torsion effect was also studied. After that, more thick brane solutions were given in Ref.~\cite{Menezes:2014bta} by the first-order formalism, and the domain wall brane in a reduced Born-Infeld-$f(T)$ theory was studied in Ref.~\cite{Yang:2017evd}. The stability of this system under the tensor perturbations was studied in Ref.~\cite{Guo:2015qbt}. In all these models, the thick branes are generated by a background scalar field. In mimetic theories, the thick brane can be generated by the mimetic scalar field. In this paper we are interested in the thick brane generated by the mimetic field in $f(T)$ theory. Compared with the matter field, the mimetic field comes from the geometry part and can generate more interesting inner structures, which will result in the following interesting characters: (1) The graviton zero mode has a deep split with a cat-like shape for some parameters, which is plotted in Fig.~\ref{figZeroModes} in order to compare with the result of the $f(T)$-brane generated by a matter field~\cite{Yang:2012hu,Guo:2015qbt}. (2) There are many gravitational resonant KK modes. Besides $f(T)=T+\alpha T^b$, we also find some solutions for other two forms of $f(T)$.

The thick brane world scenario in mimetic theory has been studied in Ref.~\cite{Zhong:2017uhn,Guo:2015qbt} recently. Some thick brane solutions in mimetic theory were found and the tensor and scalar perturbations were analysed. Before that, the late time cosmic expansion and inflation on a thin brane in mimetic gravity were investigated in Ref.~\cite{Sadeghnezhad:2017hmr}. Inspired by these works, we will investigate the brane world model in mimetic $f(T)$ theory in this paper. The stability of the brane system will be investigated by analysing the tensor perturbations of the vielbein. Besides, the effects of the torsion will also be studied by comparing different kinds of  $f(T)$.

This paper is organized as follows. In Sec.~\ref{Setup}, we will review the mimetic method and $f(T)$ gravity briefly. The Lagrange multiplier formulation of mimetic $f(T)$ theory will also be given. In Sec.~\ref{solution}, we will solve the field equations and obtain solutions of the brane system for different kinds of  $f(T)$. The stability of this system will be analysed by studying the tensor perturbations of the vielbein in Sec.~\ref{perturbation}. In this section, we will also study the localization of the zero mode of graviton on the brane, and the effects of the torsion will be given. The gravitational resonances will be analyzed in Sec.~\ref{resonance}. In the end, conclusions and discussions will be given in Sec.~\ref{conclusion}.

\section{set up}\label{Setup}
In this section, we will give a brief introduction to the mimetic $f(T)$ theory. In mimetic theories, the physical metric $g_{MN}$ is written in terms of an auxiliary metric $\tilde g_{MN}$ and a scalar field $\phi$~\cite{Chamseddine:2013kea}. So, the conformal degree of freedom can be isolated in a covariant way. We will use the capital Latin letters $M,N,P,Q,...$ label the five-dimensional coordinates, and $A,B,C,D,...$ label the tangent space coordinates. The explicit relation between the physical metric $g_{MN}$ and the auxiliary metric $\tilde g_{MN}$ is
\begin{eqnarray}\label{relationofpanda}
g_{MN}=-\tilde g_{MN}\tilde g^{PQ}\partial_P\phi\partial_Q\phi.
\end{eqnarray}
As a consequence, the scalar field satisfies the following constraint
\begin{eqnarray}\label{conditionofphi}
g^{MN}\partial_M\phi\partial_N\phi=-1.
\end{eqnarray}
It is obvious that the physical metric is invariant under the conformal transformation of the auxiliary metric as $\tilde g_{MN}\rightarrow \Omega^2(x^P)\tilde g_{MN}$, where $\Omega(x^P)$ is a function of the spacetime coordinate. The action of five-dimensional mimetic gravity is of the form
\begin{eqnarray}
S={M_*^3}\int d^5x \sqrt{-g(\tilde g_{MN},\phi)}\big[R(g_{MN}(\tilde{g}_{MN},\phi))+\mathcal L_m \big],
\end{eqnarray}
where $M_*$ is the five-dimensional mass scale and $\mathcal L_m$ is the Lagrangian of the matter fields.

The gravitational field equations can be obtained by varying the action with respect to the physical metric $g_{MN}$.
However, we must pay attention to this process. Because the physical metric $g_{MN}$ can be rewritten in terms of the auxiliary metric $\tilde g_{MN}$ and scalar field $\phi$ as Eq.~\eqref{relationofpanda}, so the variation of the physical metric is not independent.
On the other hand, the action can also be written in the Lagrange multiplier formulation equivalently
\begin{eqnarray}
S={M_*^3}\int d^5x\sqrt{-g}\big[R+\lambda(g^{MN}\partial_M\phi\partial_N\phi+1)+\mathcal L_m\big].
\end{eqnarray}
The constraint~\eqref{conditionofphi} can be gotten by varying this action with respect to the Lagrange multiplier $\lambda$.

Recently, the authors of Ref.~\cite{Mirza:2017afs} applied the mimetic method into $f(T)$ gravity, which is a modification of TEGR. In $f(T)$ gravity, the background spacetime is not a Riemann manifold anymore but the so-called Weitzenb\"{o}ck manifold. The dynamical field is the vielbein $e_A(x^M)$ defined in the tangent space at any point $x^M$ of the manifold, rather than the metric $g_{MN}$. The relation between the metric and the vielbein is $g_{MN}=\eta_{AB}{e^A}_M{e^B}_N,$
where $\eta_{AB}$ is the Minkowski metric of the tangent space with the form of $\eta_{AB}=\text{diag}(-1,1,1,1,1)$, and ${e^A}_M$ is the component of the vielbein $e_A$ in the spacetime coordinate $x^M$. From this relation, we can get
${e_A}^{M}{e^A}_{N}=\delta^{M}_{N},$ and ${e_A}^M {e^B}_M=\delta^{B}_{A}.$
The curvatureless  Weitzenb\"{o}ck connection $\tilde{\Gamma}^{P}_{~MN}$ is defined as
$\tilde{\Gamma}^{P}_{~MN}\equiv {e_A}^P\partial_N {e^A}_M,$
which can be used to construct the asymmetric  torsion tensor
${T ^{P}}_{~MN}=\tilde{\Gamma}^{P}_{~NM}-\tilde{\Gamma}^{P}_{~MN}.$
The Ricci scalar $\tilde R$ constructed by the Weitzenb\"{o}ck connection is zero, so this manifold is curvatureless.
The contorsion tensor ${K^P}_{MN}$ is defined as the difference between the Weitzenb\"{o}ck connection $\tilde{\Gamma}^{P}_{~MN}$ and the Levi-Civita connection $\Gamma^{P}_{~MN}$ which can be described as $ K^{P}_{~MN}=\frac{1}{2}\left(T^{~~P}_{M~N}+T_{N~M}^{~~P}-T^{P}_{~MN}\right).$ Besides, by defining the superpotential as $S_{P}^{~MN}\equiv\frac{1}{2}
   \left({K^{MN}}_{P}
   -{\delta^{N}_{P}{T^{QM}}_{Q}}
   +\delta^{M}_{P}{T^{QN}}_{Q}\right),$
 we can construct the torsion scalar $T=S_{P}^{~MN}T ^{P}_{~MN}.$
After some cumbersome but simple algebra, we can get that $T=-\bar R-2\bar\nabla^M {T^P}_{MP}$, where $\bar R$ and $\bar\nabla$ are the Ricci scalar and covariant derivative constructed from the Levi-Civita connection, respectively. As the difference between $\bar R$ and $T$ is only a boundary term, the teleparallel gravity is equivalent to GR.

In order to apply the mimetic method into $f(T)$ gravity, the equivalent form of Eq.~\eqref{relationofpanda} must be given. The authors of Ref.~\cite{Mirza:2017afs} kept the vielbein unchanged but changed the Minkowski metric $\eta_{AB}$ as $\eta_{AB}=-\tilde\eta_{AB}\tilde\eta^{CD}\partial_C\phi\partial_D\phi.$
The auxiliary metric is defined as $\tilde g_{MN}=\tilde\eta_{AB}{e^A}_M{e^B}_N.$
So Eq.~\eqref{relationofpanda} can be derived directly \footnote{
The physical metric is
$
g_{MN}=\eta_{AB}{e^A}_M{e^B}_N
      =-\tilde\eta_{AB}\tilde\eta^{CD}\partial_C\phi\partial_D\phi{e^A}_M{e^B}_N
      =-\tilde g_{MN}\tilde\eta^{CD}\partial_C\phi\partial_D\phi
      =-\tilde g_{MN}\tilde\eta^{CD}{e_C}^P{e_D}^Q\partial_P\phi\partial_Q\phi
      =-\tilde g_{MN}\tilde g^{PQ}\partial_P\phi\partial_Q\phi,
$ where we have used that ${e_C}^P$ is the projection from the spacetime coordinate to the  tangent space coordinate.}.
The explicit relation between the variation of the metric $g_{MN}$ and that of the vielbein was derived in Ref.~\cite{Mirza:2017afs}. Of course, there is also an equivalent Lagrange multiplier formulation of the mimetic $f(T)$ gravity. In this paper, we take the action as the form of
\begin{eqnarray}\label{actionofbrane}
S=M_*^3\int d^5x\, e\left[-\frac{1}{4}f(T)+L_\phi\right],
\end{eqnarray}
where $L_\phi=\lambda\left(g^{MN}\partial_M\phi\partial_N\phi-U(\phi)\right)-V(\phi)$, and $\lambda$ is the Lagrange multiplier. The original $-1$ in Eq.~\eqref{conditionofphi} of mimetic gravity was generalized to $U(\phi)<0$ in Ref.~\cite{Astashenok:2015haa}. It was also adopted into the brane world model in Ref.~\cite{Zhong:2017uhn} with the condition $U(\phi)>0$ since the mimetic scalar field $\phi$ generating the brane only depends on the extra dimension $y$, i.e., $g^{MN}\partial_M\phi\partial_N\phi=(\partial_y \phi(y))^2>0$~\cite{Zhong:2017uhn}. Note that, there are $\frac{D(D-3)}{2}+D$ degrees of freedom in mimetic $f(T)$ theory where $D$ is the dimension of spacetime. And one of these degrees of freedom comes from the mimetic approach, and it is free of ghost on condition that the energy density is positive \cite{Barvinsky:2013mea}. However, the $D-1$ extra degrees of freedom from $f(T)$ theory could
superluminal propagate, and there could be closed causal curves \cite{Ong:2013qja,Izumi:2013dca}.

In thick brane world model, the static flat brane metric which keeps the four-dimensional Poinc\'{a}re invariance is given by
\begin{eqnarray}
ds^2=e^{2A(y)}\eta_{\mu\nu}dx^\mu dx^\nu+dy^2,
\end{eqnarray}
where Greek letters $\mu,\nu,...$ denote the coordinates on the brane. Here $e^{2A(y)}$ is the so-called warp factor. Straightforwardly, we can choose the vielbein as the form of ${e^A}_M=\text{diag}(e^{A},e^{A},e^{A},e^{A},1)$, which has been proved to be a good choice~\cite{Tamanini:2012hg,Ferraro:2011us}. With this metric ansatz, we can express the explicit equations of motion:
\begin{eqnarray}
\frac{1}{4} \left[6 f_T A''+24 A'^2 \left(f_T-6 f_{TT} A''\right)+f(T)\right]\nn\\
+V(\phi )+\lambda \left(U(\phi)-\phi'^2\right)&=&0,\\ \label{eqofm1}
6 f_T A'^2+\frac{1}{4}f(T)+V(\phi )+\lambda  \left(U(\phi)+\phi '^2\right)&=&0,\\ \label{eqofm2}
\lambda\big(8 A'\phi '+\frac{d U(\phi)}{d \phi}+2 \phi ''\big)+\frac{d V(\phi )}{d \phi}+2 \lambda ' \phi '&=&0,\\ \label{eqofm3}
\phi '^2-U(\phi)&=&0. \label{eqofm4}
\end{eqnarray}
Here, we denote the primes as the derivatives with respect to the extra dimension $y$. Note that there are three independent equations only. The Lagrange multiplier $\lambda$ can be solved by substituting Eq.~\eqref{eqofm4} into Eq.~\eqref{eqofm1}. Next, we will choose different kinds of  $f(T)$ to analyse the effect of torsion on the brane world.

\section{Solutions}\label{solution}

We need to solve five functions: $A(y)$, $\phi(y)$, $\lambda(y)$, $U(\phi)$, and $V(\phi)$, but there are only three independent equations. So we start by giving the warp factor $A(y)$ and the mimetic scalar field $\phi(y)$. We will consider two kinds of warp factors in this section and give solutions of the mimetic $f(T)$ brane world model.

\begin{figure}[htb]
\begin{center}
\subfigure[~The warp factor~(\ref{warp1})]  {\label{warpfactor1}
\includegraphics[width=4cm]{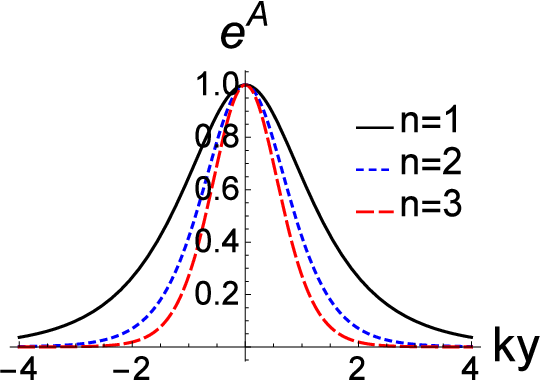}}
\subfigure[~The warp factor~(\ref{warp2})]  {\label{warpfactor2}
\includegraphics[width=4cm]{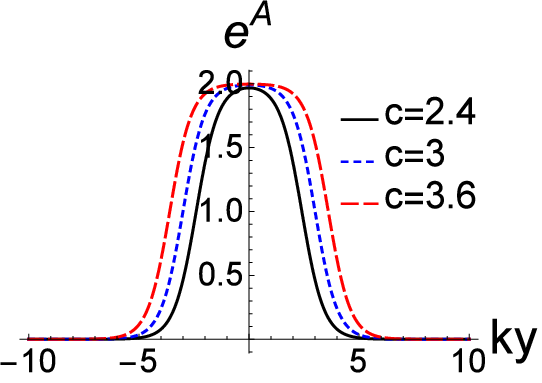}}
\subfigure[~The scalar field~(\ref{scalarfield})] {\label{phi}
\includegraphics[width=4cm]{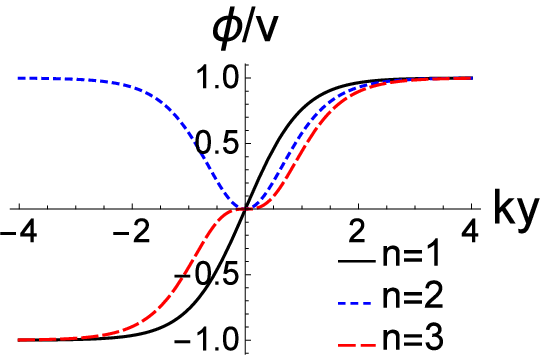}}
\end{center}
\caption{Plots of the warp factors~(\ref{warp1}) and (\ref{warp2}) and the scalar field~(\ref{scalarfield}).}
\label{warpphi}
\end{figure}

We consider the mimetic scalar field  as
\begin{eqnarray}
\phi(y)=v\tanh ^n(k y),\label{scalarfield}
\end{eqnarray}
where the parameter $k$ has mass dimension one, $v$ is a positive parameter which representing the limit of the scalar field. If $n$ is odd, the scalar field is a single-kink (the black line in Fig.~\ref{phi}) or a double-kink (the red line in Fig.~\ref{phi}). If $n$ is even, the scalar field is not a kink configuration (the blue line in Fig.~\ref{phi}). Usually, the formation of a thick brane requires a kink configuration scalar field, but in mimetic theory, the non-kink scalar field can also generate a thick brane, because of the Lagrange multiplier. We take two kinds of warp factors,
\begin{eqnarray}
A(y)&=&-n \ln (\cosh (k y)), \label{warp1} \\
A(y)&=&\ln[\tanh (k (y + c))-\tanh (k (y- c))].\label{warp2}
\end{eqnarray}
The shapes of these two kinds of warp factors and the mimetic scalar field are shown in Fig.~\ref{warpphi}. From Fig.~\ref{warpfactor2} we can see that the warp factor~(\ref{warp2}) has a platform near the origin of the extra dimension. We can adjust the width of the platform by changing the parameter $c$.
When $y\rightarrow\pm \infty$, $A(y)\rightarrow-nk|y|$, so the spacetime is asymptotically $\text{AdS}_5$.

Then we will give solutions for different kinds of  $f(T)$ for the warp factor~(\ref{warp1}). We do not show solutions for the warp factor~(\ref{warp2}), because they are complicated and tedious. From Eq.~\eqref{eqofm4} we know that $U(\phi)$ depends on $\phi$ only, so it can be solved as
\begin{eqnarray}
U(\phi)=k^2n^2v^2\Phi^{\frac{2(n-1)}{n}}\left(\Phi^{\frac{2}{n}}-1\right)^2,
\end{eqnarray}
where $\Phi={\phi}/{v}$  and the parameter $v$ is the same one that appears in Eq.~(\ref{scalarfield}).
For $f(T)=T$, $f(T)$ theory degenerates to TEGR. The solution of this model was given in Ref.~\cite{Zhong:2017uhn}:
\begin{eqnarray}
\lambda(y)&=&-\frac{3}{2nv^2}\sinh^2(ky)\tanh^{-2n}(ky),\\
V(\phi)&=&3k^2\left[n-n(1+2n)\Phi^{\frac{2}{n}}\right].
\end{eqnarray}

The second kind of $f(T)$ is taken as $f(T)=T+\alpha T^b$, where the mass dimension of the parameter $\alpha$ is $2-2b$ to ensure that $\alpha T^b$ is mass dimension $2$. The solution is
\begin{eqnarray}
\lambda(y)&=&\alpha  12^b b \frac{12 k^2 n^2 \cosh ^2(k y)\tanh ^2(k y)}{16 k^2 n^3 v^2  \tanh ^{2 n}(k y)}\nn\\
           &\times& (2 b-1)\cosh ^2(k y)\left(k^2 n^2 \tanh ^2(k y)\right)^b,\\
V(\phi)&=&\frac{3}{2}k^2 n \big(1-(2 n+1) \Phi^{\frac{2}{n}}\big)+\frac{3}{2}k\alpha(2 b-1)\nn \\
          &\times& 12^{b-1}(kn)^{2b-1} \big((b+2 n) \Phi^{\frac{2}{n}}-b\big)   \Phi^{\frac{2b-2}{n}}.
\end{eqnarray}

As for $f(T)=T_0\tanh\left(\frac{T}{T_0}\right)$ and $f(T)=T_0\, e^{\frac{T}{T_0}}$, the solutions are tedious, so we do not show them.

Next, we will analyse the stability of this system by investigating tensor perturbations of the vielbein. Besides, the effects of different kinds of  $f(T)$ will also be investigated.

\section{Tensor perturbations and Localization}\label{perturbation}
In this section, we will consider the tensor perturbations of this mimetic $f(T)$ brane world model. The perturbed four-dimensional vielbein is ${e^a}_\mu=e^{A(y)}({\delta^a}_\mu +{h^a}_\mu)$~\cite{Guo:2015qbt}
where the Latin letters $a, b,...$ denote the tangent space coordinates on the brane. Note that, Eq. (33) in Ref.~\cite{Mirza:2017afs} give the explicit formulation of the variation of the physical metric $g_{MN}$. For tensor perturbations considered here, we do not need to consider the perturbation of the scalar field $\phi$ since they are decoupled. So we can get the perturbed four-dimensional physical metric $g_{\mu\nu}=e^{2A(y)}(\eta_{\mu\nu}+\gamma_{\mu\nu})$,
where
$\gamma_{\mu\nu}=({\delta^a}_\mu {h^b}_\nu+{\delta^b}_\nu {h^a}_\mu)\eta_{ab}.$
We impose the transverse and traceless condition that $\partial_\mu\gamma^{\mu\nu}=0=\eta^{\mu\nu}\gamma_{\mu\nu}$, whose equivalent vielbein form is
$\partial_\mu({\delta_a}^\mu {h_b}^\nu+{\delta_b}^\nu{ h_a}^\mu)\eta^{ab}={\delta_a}^\mu {h^a}_\mu=0.$
Substituting the perturbed vielbein into the equation of motion, after some cumbersome but simple caculations, we can get the perturbed field equation:
\begin{eqnarray}
\left(e^{-2A}\square^{(4)}\gamma_{\mu\nu}+\gamma''_{\mu\nu}+4A'\gamma'_{\mu\nu}\right)f_T\nn\\ -24A'A''\gamma'_{\mu\nu}f_{TT}=0,
\label{mainequation}
\end{eqnarray}
where $\square^{(4)}=\eta^{\mu\nu}\partial_\mu\partial_\nu$ is the four-dimensional d'Alembert operator. Making the coordinate transformation $dz=e^{-A}dy,$
Eq.~\eqref{mainequation} becomes to
\begin{equation}
\left(\partial_z^2+2H\partial_z+\square^{(4)}\right)\gamma_{\mu\nu}=0, \label{perturbationEq2}
\end{equation}
where
$H=\frac{3}{2}\partial_z A
       +12e^{-2A}\left(\left(\partial_z A \right)^3
                      -\partial_z^2 A\partial_z A
               \right)
        \frac{f_{TT}}{f_T}.$
Then we introduce the KK decomposition
$
\gamma_{\mu\nu}(x^\rho,z)=\epsilon_{\mu\nu}(x^\rho)e^{\int{-H(z)dz}} \psi(z).
$
After that, we can get a four-dimensional Klein-Gordon-like equation for the four dimensional graviton $\epsilon_{\mu\nu}$: $\left(\Box^{(4)}+m^2\right)\epsilon_{\mu\nu}(x^\rho)= 0$
and a Schr\"odinger-like equation for the extra-dimensional profile $\psi(z)$:
\begin{eqnarray}
 \left(-\partial_z^2+W(z)\right)\psi = m^2\psi, \label{SchrodingerEquation}
\end{eqnarray}
where $W(z)=H^2+\partial_z H$ is the effective potential. This Schr\"odinger-like equation can be factorized into a supersymmetric quantum mechanics form: $\big(\partial_z+H\big)\big(-\partial_z+H\big)\psi=m^2\psi,$
which guarantees that all the eigenvalues $m^2$ are non-negative, that is to say, there is no tachyonic graviton. So this model is stable under tensor perturbations with the transverse-traceless condition. Of course, there is a zero mode $m=0$ for
this system, and the solution is $\psi_0=N_0e^{\int{-H(z)dz}}$,
where $N_0$ is the normalization coefficient.

In order to recover the four-dimensional gravity, the zero mode of graviton should be localized near the origin of the extra dimension. This requires that $\int dz \;\psi_0^2(z) < \infty$. We will investigate this issue in this section.

\begin{figure}[htb]
\begin{center}
\subfigure[~The effective potential]  {\label{effective potential1}
\includegraphics[width=4cm]{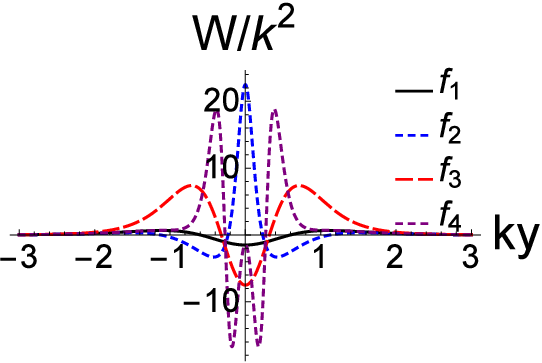}}
\subfigure[~The zero mode] {\label{psi1}
\includegraphics[width=4cm]{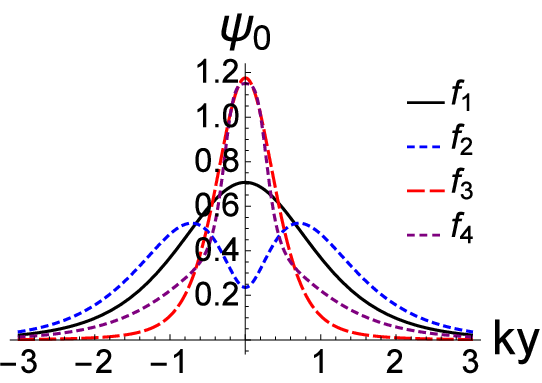}}
\end{center}
\caption{Plots of the effective potential and the zero mode for $A(y)=-n \ln (\cosh (k y))$. The parameter is set to $n=1$. The symbols $f_i$ in the figures denote $f_1=T$, $f_2=T+\alpha T^b$, $f_3=T_0e^{\frac{T}{T_0}}$, and $f_4=T_0\tanh\left(\frac{T}{T_0}\right)$ with $T_0=1$.}
\label{Upsi1}
\end{figure}

As for the analytical formulations of the effective potential and the zero mode for some solutions are complicated, we just give that of $f(T)=T$, and for the other cases we only show their plots.
\begin{eqnarray}
W(y)&=&\frac{3k^2 n  [5 n \cosh (2 k y)-5 n-4]}{8\cosh ^{2 (n+1)}(k y)} ,\\
\psi_0(y)&=&N_0\cosh ^{-\frac{3 n}{2}}(k y).
\end{eqnarray}
The normalization coefficient $N_0$ can be fixed by $\int^{+\infty}_{-\infty}\psi_0^2(z)dz=\int_{-\infty }^{+\infty } e^{-A(y)}\psi_0^2(y)dy=1$. In this case it can be calculated as $N_0=\frac{\sqrt{\pi } \Gamma (n)}{k \Gamma \left(n+\frac{1}{2}\right)}$ with $\Gamma(n)$ the Gamma function. As shown in Fig.~\ref{Upsi1} (the black lines), we can see that the  effective potential is volcano-like and the corresponding zero mode has only one peak and it tends to zero at infinity. Besides, the four-dimensional gravity can be recovered. This is expected because when $f(T)=T$, $f(T)$ gravity is equivalent to GR and this result has been obtained in~\cite{Zhong:2017uhn}.

The effective potentials and the zero modes are shown in Fig.~\ref{Upsi1}. From this figure we can see that for $f(T)=T+\alpha T^b$ (the blue lines in Fig.~\ref{Upsi1}) there is a double well in the effective potential and a split in the zero mode, though the warp factor does not split. For $f(T)=T_0e^{\frac{T}{T_0}}$ (the red lines in Fig.~\ref{Upsi1}), the shapes of the effective potential and the zero mode are similar to that of $f(T)=T$, but the well of effective potential is deeper, and the corresponding zero mode is sharper. The purple lines in Fig.~\ref{Upsi1} are corresponding to $f(T)=T_0\tanh\left(\frac{T}{T_0}\right)$. We can see that though the effective potential has two sub-wells, the zero mode does not split, because the sub-wells are too narrow.

\begin{figure}[htb]
\begin{center}
\subfigure[~The effective potential]  {\label{effective potential21}
\includegraphics[width=4cm]{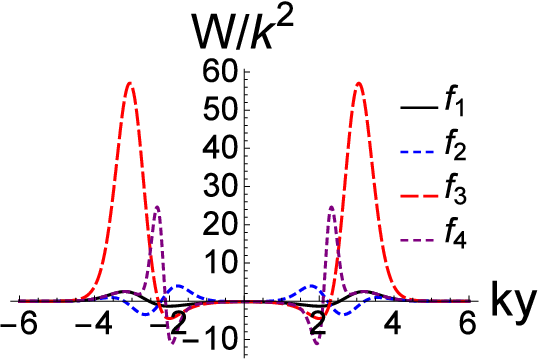}}
\subfigure[~The zero mode] {\label{psi2}
\includegraphics[width=4cm]{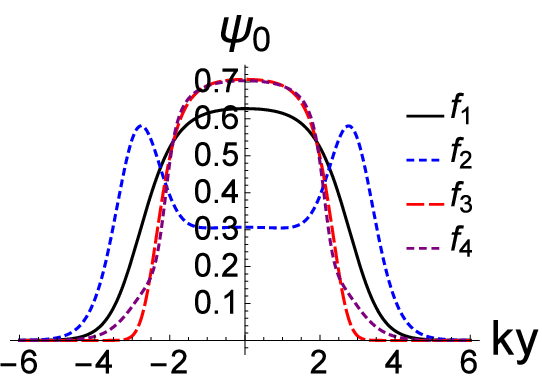}}
\end{center}
\caption{Plots of the effective potential and the zero mode for $A(y)=\ln[\tanh (k (y +c))-\tanh (k (y-c))]$. The parameter is set to $c=1$. The symbols $f_i$ are the same as Fig.~\ref{Upsi1}.}
\label{Upsi2}
\end{figure}

For the solutions of the warp factor~(\ref{warp2}), the effective potentials and the zero modes are shown as in Fig.~\ref{Upsi2}. From this figure we can see that for $f(T)=T$ (the black line) and $f(T)=T_0e^{\frac{T}{T_0}}$ (the red line) there is a platform in the zero mode which is corresponding to the warp factor (see Fig.~\ref{warpfactor2}). Comparing the polynomial kind of $f(T)$ with $f(T)=T$, we can see that there is also a split in the zero mode based on the platform, we call it the cat-like mode. So we can conclude that if $f(T)$ takes the polynomial form the zero mode of graviton may have a split. For $f(T)=T_0\tanh\left(\frac{T}{T_0}\right)$ (the purple lines in Fig.~\ref{Upsi2}), the effective potential and the zero mode are the same as the case of for $f(T)=T$. Besides, we have calculated that $\int^{+\infty}_{-\infty}\psi_0^2(z)dz$ is converged, so they can be localized near the brane for all the three models.

\begin{figure}[htb]
\begin{center}
\subfigure[~The effective potential]  {\label{effective potentialn}
\includegraphics[width=4cm]{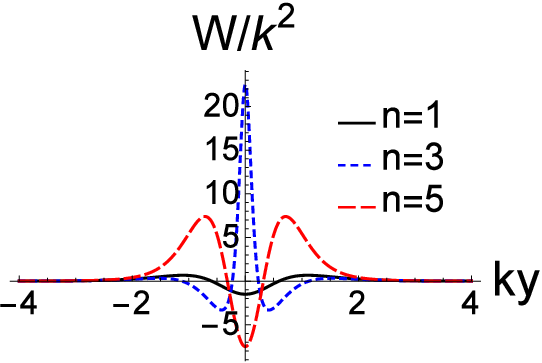}}
\subfigure[~The zero mode] {\label{psin}
\includegraphics[width=4cm]{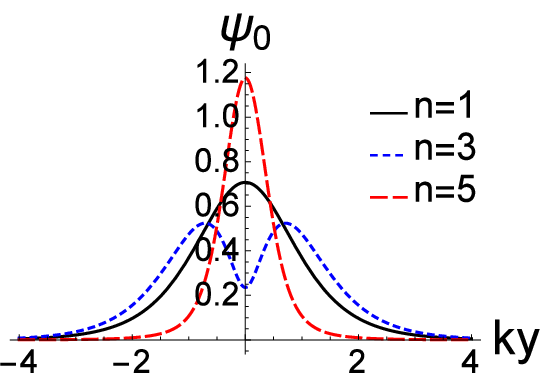}}
\end{center}
\caption{Plots of the effective potential and the zero mode of $f(T)=T$ for the warp factor~(\ref{warp1}).}
\label{Upsin}
\end{figure}

\begin{figure}[htb]
\begin{center}
\subfigure[~The effective potential]  {\label{effective potentialb}
\includegraphics[width=4cm]{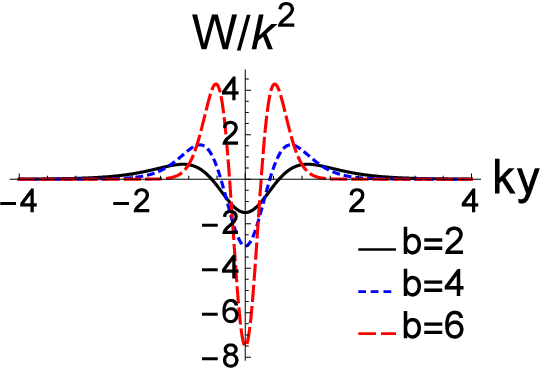}}
\subfigure[~The zero mode] {\label{psib}
\includegraphics[width=4cm]{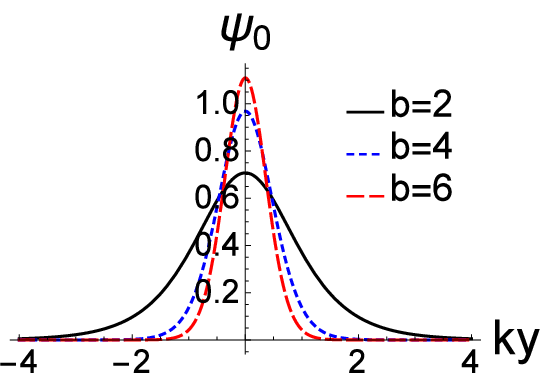}}
\end{center}
\caption{Plots of the effective potentials and the zero modes of $f(T)=T+\alpha T^2$ for the warp factor~(\ref{warp1}). The parameters are set to $n=1$, $\alpha=-1$.}
\label{Upsib}
\end{figure}

We take $f(T)=T$ as an example to analyse the effects of the parameter $n$ in the warp factor~(\ref{warp1}) on the effective potential and the zero mode.  We can conclude from Fig.~\ref{Upsin} that the depth of the effective potential increases with the parameter $n$. As a result, the corresponding zero mode becomes sharper. As for the parameter $b$ characterizing the polynomial kind of $f(T)$, the effects can be seen from Fig.~\ref{Upsib}. Note that, the two curves with black lines in Figs.~\ref{effective potential1} and~\ref{psi1} are the same as those with blue lines in Figs.~\ref{effective potentialb} and~\ref{psib}, respectively. It can be seen that there are two barriers in the effective potential when $b>2$ and the zero modes are all split for $b>1$. Besides, the heights of the barrier of the effective potential and the normalized  zero mode increase and decrease with the parameter $b$, respectively. We note that the zero mode does not localize near the origin of the extra dimension, but localize near the ``ears" of the cat-like energy density.

\section{Gravitational resonant}\label{resonance}
In order to study the graviton resonances, Almeida et al. proposed the large peaks of the wavefunction as the resonances~\cite{Almeida:2009jc}. The transfer matrix method was also use to find the resonances \cite{Landim:2011ki,Du:2013bx}. We use the relative possibility $P$ proposed in Ref.~\cite{Liu:2009ve}: $P(m^2)=\frac{\int_{-z_b}^{z_b}|\psi(z)|^2dz}{\int_{-z_{max}}^{z_{max}}|\psi(z)|^2dz}$, where $2z_b$ can be regarded as the width of the brane approximately, $z_{max}=10z_b$, and $\psi(z)$ is the solution of the Schr\"odinger-like equation \eqref{SchrodingerEquation}. Besides, note that the effective potentials are even functions, so the wave functions will be either even or odd. So we can consider $\psi_{\text{even}}(0)=1, \partial_z\psi_{\text{even}}(0)=0$ for even parity and $\psi_{\text{odd}}(0)=0, \partial_z\psi_{\text{odd}}(0)=1$ for odd parity to solve the Schr\"odinger-like equation \eqref{SchrodingerEquation} numerically \cite{Liu:2009ve}.

\begin{figure}[htb]
\begin{center}
\subfigure[~The relative probability]  {\label{relative probability}
\includegraphics[width=4cm]{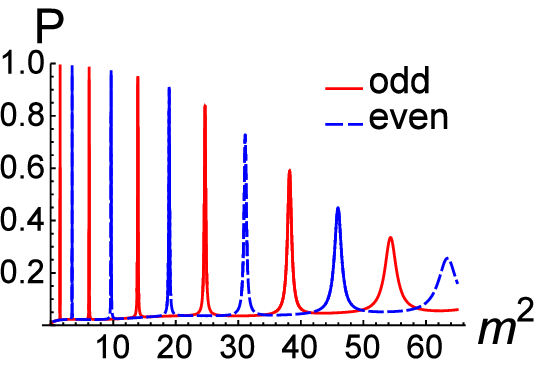}}
\subfigure[~The resonances] {\label{eigenfunction}
\includegraphics[width=4cm]{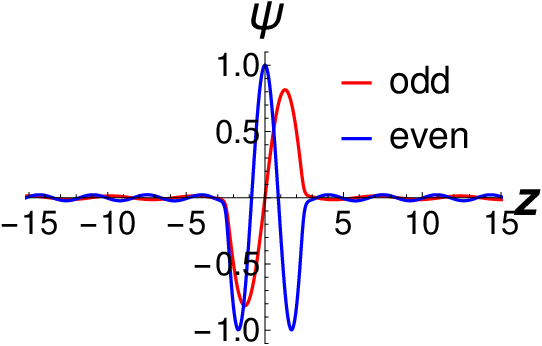}}
\end{center}
\caption{Plots of the relative probability and the first two resonances for $f(T)=T-2T^4$. The parameter $c$ is chosen to be $6$.}
\label{Pf}
\end{figure}

We do not find resonance for solutions with the warp factor~(\ref{warp1}), so we will just consider the solutions with the warp factor~(\ref{warp2}). From Fig.~\ref{effective potential21} we can see that all these four kinds of forms of $f(T)$ can support gravitational resonant modes. The number of gravitational resonance modes increases the width and the height of the barrier of the effective potential. So different parameters have different effects on the effective potential and the gravitational resonance modes.  As the width of the potential barriers increases with the parameter $c$, so the number of the graviton resonance increases with the value of parameter $c$. For the second form of $f(T)$, the effects of torsion are caused by the second term. The height of the barrier of the effective potential increases with parameter $b$, and decreases with the absolute value of $\alpha$. So the number of the graviton resonance increases with parameter $b$, and decreases with the absolute value of $\alpha$. For $f(T)=T_0 e^{\frac{T}{T_0}}$ and $f(T)=T_0 \tanh\left(\frac{T}{T_0}\right)$, the parameter $T_0$ only affects the height of the effective potential barrier. And the height decreases with the parameter $T_0$, so the number of the gravitational resonance modes decreases with the parameter $T_0$. A specific example of $f(T)=T-2T^4$ with $c=6$ is shown in Fig.~\ref{Pf}.

\section{Conclusions and discussions}\label{conclusion}

In this paper, we applied the mimetic $f(T)$ theory into the brane world model. The brane is supported by the isolated conformal degree of freedom, i.e., the mimetic scalar field. We used the Lagrange multiplier formulation and derived the equations of motion. By taking the flat static brane metric which reserves the four-dimensional Poincare invariance, we obtained the explicit equations of motion. We found that the non-kink configuration of the scalar field can also generate a thick brane. The warp factor has two forms: $A(y)=-n \ln (\cosh (k y))$ and $A(y)=\ln[\tanh (k (y + c))-\tanh (k (y- c))]$. For both cases, we found solutions for four kinds of $f(T)$. For $f(T)=T+\alpha T^b$, we found the cat-like thick brane for some specific parameters.

By investigating the tensor perturbations of the vielbein, we analysed the stability of this system. It was found that the perturbed field equation can be transformed to a Klein-Gordon-like equation and a Schr\"odinger-like equation after the KK decomposition. The Schr\"odinger-like equation can further be factorized into a supersymmetric quantum mechanics form, which guarantees that there are no tachyonic gravitons. Therefore, the brane in mimetic $f(T)$ theory is stable under the tensor perturbations. We also studied the localization of gravity and found that for all the solutions the four-dimensional gravity can be recovered.

As we can see from Fig.~\ref{effective potential21}, the potential wells are deep enough, so there are many gravitational resonant KK modes for the solutions with the warp factor~(\ref{warp2}). We analysed the effects of different parameters and showed the plot of the gravitational resonant KK modes for $f(T)=T+\alpha T^2$.  All these gravitational resonant KK modes contribute to the  four-dimensional Newtonian potential, so they can cause different results compared with that of $f(T)$-brane generated by a matter field~\cite{Yang:2012hu}. As pointed out in Ref.~\cite{Zhong:2017uhn}, scalar perturbations do not propagate on the brane in mimetic gravity, which is different from the case of brane world based on GR. Therefore, it is worth to investigate scalar perturbations of the brane system in mimetic $f(T)$ gravity. Moreover, thick branes with inner structure generated by more than one mimetic scalar fields are also interesting.

\section{Acknowledgments}

This work was supported by the National Natural Science Foundation of China (Grants Nos. 11875151 and 11522541), and the Fundamental Research Funds for the Central Universities (Grants No. lzujbky-2019-it21 and lzujbky-2018-k11).

%

%
%
%

\end{document}